\documentclass[prd,preprint,superscriptaddress,showkeys]{revtex4}
\usepackage{revsymb}
\usepackage{amsmath}
\newcommand{\anf}{``}
\begin{document}
\title{INTERACTING DARK ENERGY AND COSMOLOGICAL EQUATIONS OF STATE}
\author{Winfried Zimdahl\footnote{E-mail address: zimdahl@thp.uni-koeln.de}}
\address{Institut f\"ur Theoretische Physik, Universit\"at zu  K\"oln\\
D-50937 K\"oln, Germany}

\begin{abstract}
Interactions within the cosmic medium modify its equation of
state. We discuss implications of interacting dark energy models
both for the spatially homogenous background and for the
perturbation dynamics.
\end{abstract}

\maketitle

\section{Introduction}

According to our current understanding the presently observable
Universe is dynamically dominated by two so far unknown
substances, Dark Matter (DM) and Dark Energy (DE). Of major
interest is a precise knowledge of the DE equation of state (EOS)
including its possible time dependence. A definitely time
dependent EOS would rule out a ``true" cosmological constant which
still is the most favored DE candidate. A related problem is to
understand why the energy densities of both components are of the
same order of magnitude today (the ``coincidence problem").

Most approaches in the field rely on an independent evolution of
DE and DM. Given the unknown nature of both DE and DM one may
argue, however, that an entirely independent behavior is a very
special case \cite{Amendola,plb}. At least, there is no compelling
reason  to exclude interactions from the outset. Unified models
such as the Chaplygin gas models even try to understand DE and DM
as different manifestations of one single fluid \cite{Kamen}.

Our aim here is to point out the role of interactions within the
dark sector for the cosmological dynamics. As far as the
coincidence problem is concerned, an interaction may naturally
lead to a fixed ratio of the energy densities of DM and DE. An
interaction will also influence the perturbation dynamics. In
particular, it modifies the adiabatic sound speed and the large
scale perturbation behavior. It affects the time dependence of the
gravitational potential and hence the Integrated Sachs-Wolfe
effect (ISW) which is observable through the lowest multipoles of
the anisotropy spectrum of the cosmic microwave background.
Furthermore, a coupling between DM and DE is expected to be
relevant for holographically determined DE and for the big rip
scenario.

\section{Interacting Dark Energy}

An interaction between DM and DE is most conveniently modelled as
a decay of the latter into the former,
\begin{equation}\label{dotrhoMX}
\dot{\rho}_{M} + 3H \rho_{M}= \Gamma \rho_{X}\ ,\qquad
\dot{\rho}_{X} + 3H \left(1 + w_X\right)\rho_{X}= -\Gamma
\rho_{X}\ ,
\end{equation}
where $\rho_{M}$ and $\rho_{X}$ are the energy densities of DM and
DE, respectively, $H$ is the Hubble expansion rate and $w_{X}$ is
the EOS parameter of the DE. The (not necessarily constant) decay
rate is denoted by $\Gamma$. Einstein's field equations for
spatially flat FRLW cosmologies are
\\
\begin{equation}
H^{2} = \frac{8\, \pi G}{3} \rho \ ,  \qquad \dot{H}  = - 4\, \pi
G (\rho + p_X) \, , \label{efe}
\end{equation}
where $\rho = \rho_M + \rho_X$ is the total energy density. The
quantity $\dot H$ is related to the deceleration parameter $q$ by
$q =- 1-\frac{\dot H}{H^2}$. It is obvious, that neither the
Hubble parameter $H$ nor the deceleration parameter is directly
affected by the interaction (there is an indirect influence,
however, since the coupling modifies the ratio
$\frac{\rho_{M}}{\rho_{X}}$ ). It is only the second derivative of
the Hubble parameter, equivalent to the third derivative of the
scale factor $a$ (where $H \equiv \frac{\dot{a}}{a}$) in which the
decay rate enters explicitly,
\begin{equation}
\frac{\ddot{H}}{H^{3}} = \frac{9}{2} +
\frac{9}{2}w_{X}\frac{\rho_X}{\rho}\left[2 + w_X +
\frac{1}{3H}\left(\Gamma - \frac{\dot{w}_X}{w_X}\right)\right]\ .
\label{doubledotH}
\end{equation}
This corresponds to the fact that $\Gamma$ enters the luminosity
distance $d_L = (1+z)\int \frac{dz}{H}$ only in third order of the
redshift $z$ (cf.~Refs.~\cite{WDScale,Marek}). Consequently, the
interaction becomes observable if the cosmological dynamics is not
only described by the present values $H_0$ and $q_0$ of $H$ and
$q$, respectively, but additionally by parameters that involve at
least $\ddot{H}_{0}$, the present value of $\ddot{H}$. The
recently introduced \anf statefinder" parameters \cite{sahni,alam}
are of this type and represent useful tools to characterize
interacting models \cite{WDGRG}.

\section{Constant Density Ratio}

Assuming $w_X =$ const there exists a special solution of the
system (\ref{dotrhoMX}) for which
$\kappa\equiv\frac{\rho_{M}}{\rho_{X}} = \kappa_{0}=$ const. This
solution which is of interest for the coincidence problem
\cite{plb} requires a decay rate
\begin{equation}
\Gamma = - 3\,H\,\frac{\kappa_{0}}{1 + \kappa_{0}}\, w_X \
.\label{Gamma}
\end{equation}
The EOS parameter $w_X$ has to be negative to obtain $\Gamma
> 0$, i.e., a decay of the
component $X$ into the matter component. Under this condition the
energy densities scale as
\begin{equation}
\rho_{M},\, \rho_{X},\,  \rho \propto a^{-3\left(1 + w\right)}
,\qquad w = \frac{w_X}{1 + \kappa_{0}} \ .\label{rhoprop}
\end{equation}
Here, $w$ is the {\it overall} EOS which, compared with $w_X$,  is
reduced in magnitude by a factor $1 + \kappa_{0}$. The time
dependence of the scale factor in the spatially flat case is $a
\propto t^{\frac{2}{3\left(1 + w\right)}}$.

In a next step we realize that the chain of arguments leading to
(\ref{rhoprop}) (via (\ref{dotrhoMX}) and (\ref{Gamma})) may be
reversed. The resulting statement is that any cosmic EOS $w$ can
be interpreted as an effective EOS of an interacting mixture of a
component with EOS $w_{X}$ and a dust component (a generalization
to an arbitrary constant EOS is possible). With other words, for
any one-component description of the cosmic medium there exists an
equivalent interacting two-component model which generates the
same background dynamics. This equivalence (which implies a
degeneracy with respect to observational results) will turn out to
be useful on the perturbative level. It provides us with an
internal structure for any one-component cosmic medium and
represents a systematic way to generate non-adiabatic
perturbations which we shall consider later on.

An example of how a solution $\kappa_{0}= $ const may be
approached as the result of a dynamical evolution of the energy
density ratio $\kappa$ can be sketched as follows \cite{iqs}. For
a decay rate $\Gamma = 3Hb^{2}\left(1 + \kappa\right)$ with a
parameter $b^{2} =$ const, the set of equations (\ref{dotrhoMX})
admits two stationary solutions, $\kappa_{+}$ and $\kappa_{-}$,
for the ratio $\kappa$ of the energy densities. These have the
properties $\kappa_{+} - \kappa_{-} \geq 0$ and
$\kappa_{+}\kappa_{-} = 1$. For $\kappa = \kappa_{+} =
\kappa_{-}^{-1}$ we have $\rho_{M} > \rho_{X}$, i.e., matter
dominance, while for $\kappa =\kappa^{-}$ the reverse relation
$\rho_{M} < \rho_{X}$ holds, equivalent to a DE dominated phase.
There exists a dynamical solution $\kappa\left(t\right)$ according
to which the density ratio evolves from $\kappa = \kappa_{-}^{-1}$
for small values of the scale factor (matter dominance) towards a
stable, stationary solution $\kappa = \kappa_{-}$ for large values
of the scale factor (DE dominance). The deceleration parameter
changes from positive values at $\kappa_{-}^{-1}$ to negative
values at $\kappa_{-}$. This means, the interaction drives the
transition from a matter dominated era to a subsequent phase with
accelerated expansion. The solution $\kappa = \kappa_{-}$ can be
identified with the solution $\kappa_{0}$, given by
(\ref{rhoprop}) for $b^{2} = - \kappa_{0}\,w_X\, \left(1 +
\kappa_{0}\right)^{-2}$. Consequently, a stationary ratio for a
density ratio which generates accelerated expansion is obtained as
the result of a dynamical evolution.

As already mentioned, the overall equation of state $w$ in
(\ref{rhoprop}) is reduced in magnitude  compared with $w_X$. In
particular, one may ask whether it is possible to have $w \geq -1$
while $w_X < -1$. To this purpose it is useful to consider the
basic equations (\ref{dotrhoMX}) for the simple case $\Gamma =$
const. Even if at the start of the decay the evolution is
dominated by $\rho_X$, after a short time $\rho_M$ will be
approximately
\begin{equation}\label{rhoMapprox}
\rho_{M} \approx \frac{2}{3}\frac{\Gamma H^{-1}}{1 - w_X}\rho_X\ .
\end{equation}
The calculations leading to this result are similar to those at
the end of an out-of-equilibrium decay  of (effectively) dust
matter into radiation in inflationary scenarios
\cite{KolbT,MNRAS}. It is obvious, that the ratio
$\frac{\rho_{M}}{\rho_{X}}$ in (\ref{rhoMapprox}) can only be
constant for $H=$ const, i.e., for exponential expansion.
Combining now (\ref{rhoMapprox}) with (\ref{Gamma}) we obtain a
direct relation between $w_X$ and $\kappa_{0}$,
\begin{equation}\label{wXkappa}
w_{X} = - \frac{1 + \kappa_{0}}{1 - \kappa_{0}}\ .
\end{equation}
This configuration has $w_{X} < -1$ necessarily. It approaches
$-1$ only in the limit of a vanishing matter component. While a
non-interacting DE component with $w_{X} < -1$ is known to lead to
a singularity in a finite time \cite{caldwell,caldwell2} (see,
however, Ref.~\cite {McInnes}), our example demonstrates that a
suitable interaction makes the expansion de Sitter like, thus
avoiding the big rip. Similar conclusions for interacting phantom
energy have been obtained in Refs.~\cite{Guo} - \cite{Quiros}.
(Alternative settings with interactions that lead to a transition
from $w_{X}
> -1$ to $w_{X} < -1$ were discussed in Refs.~\cite{Bo} and
\cite{Odintsov}).

It is also interesting to see how a solution (\ref{rhoprop}) with
$\kappa = \kappa_{0}$ is related to the holographic bound on the
DE \cite{cohen}. With an infrared cutoff scale $L$ the DE density
is $\rho_{X}^{hol} =3c^{2}M_{P}^{2}L^{-2}$ where $M_{P}^{2}
=8\,\pi\,G$ (cf. Ref.~\cite{Li}). It seems natural to identify $L$
with the Hubble scale $H^{-1}$. With this choice it follows from
Friedmann's equation that both $\rho_{X}$ and $\rho_{M}$ are
proportional to $H^{2}$. While this property was recently used to
discard a cutoff $L = H^{-1}$ for independently evolving
components \cite{Li}, the situation is entirely different in the
context of interacting models \cite{Horvat,HolDE}. Namely, the
same dependence of both $\rho_{X}$ and $\rho_{M}$ on $H$ can be
regarded as the result of a suitable interaction between these
components in the sense described by (\ref{Gamma}) and
(\ref{rhoprop}). With a cutoff $L = H^{-1}$ we have $\kappa =
\frac{\rho_{M}}{\rho_{X}} = \kappa_0 =$ const from the outset .
Moreover, it immediately relates the parameter $c^{2}$ in the
expression for $\rho_{X}^{hol}$ to the ratio  $\kappa_0$ by
$\kappa_0 = \frac{1 - c^{2}}{c^{2}}$.  It follows that $c^2 <1$,
different from the case of non-interacting models \cite{Li} where
$L$ is identified with the future event horizon (for an
interacting model in the latter context see Ref.~\cite{Wangetal}).
The condition for accelerated expansion, $-1 <w \leq -1/3$, then
translates into
\begin{equation}\label{wbeta}
- \left(\kappa_0 + 1\right) < w_{X} < -
\frac{1}{3}\,\left(\kappa_0 + 1\right)\, \quad {\rm equivalent\,\,
to} \quad - \frac{1}{c^{2 }} < w_{X} < - \frac{1}{3 c^{2}} \ .
\end{equation}
We emphasize that the interaction is essential here to have
acceleration. There is no non-interacting limit. Without
interaction acceleration is impossible, i.e., acceleration is a
pure interaction phenomenon in this picture.

\section{Perturbations}

Adiabatic perturbations of the cosmic fluid are characterized by
$\hat{p} = \frac{\dot{p}}{\dot{\rho}}\hat{\rho}$, where $\hat{p}$
and $\hat{\rho}$ denote the total pressure and energy density
perturbations, respectively. Both quantities are related by the
adiabatic sound speed $\frac{\dot{p}}{\dot{\rho}}$. If the
substratum is understood as composed of two subcomponents with
component $X$ decaying into component $M$, the decay rate $\Gamma$
may fluctuate. Fluctuating decay rates are known to produce
curvature perturbations in certain inflationary scenarios
\cite{Dvalietal,MataRiot,Lythetal}. Here, in a different context,
they are used to characterize internal structure in the dark
sector of the cosmic substratum \cite{WJ}. For the medium
discussed here we find
\begin{equation}\label{hatphatGamma}
\hat{p} - \frac{\dot{p}}{\dot{\rho}}\hat{\rho} =
\rho\left(\frac{\Gamma}{3 H}\right)^{\hat{}}\ ,
\end{equation}
where $\left(\frac{\Gamma}{3 H}\right)^{\hat{}}$ denotes
fluctuations about the (constant) background ratio
$\frac{\Gamma}{3 H}$. A fluctuating fractional decay rate
$\frac{\Gamma}{3H}$ necessarily modifies the adiabatic sound
speed. In particular, for a fluctuation of the type
$\left(\frac{\Gamma}{3 H}\right)^{\hat{}} = \lambda
\frac{\hat{\rho}}{\rho}$ we obtain an effective sound speed square
$c_{\rm eff}^{2} = \frac{\dot{p}}{\dot{\rho}} + \lambda$. Even if
$\frac{\dot{p}}{\dot{\rho}}$ is negative, a non-adiabatic
contribution due to an internal structure may give rise to a
physically acceptable effective sound speed. On the perturbative
level the cosmic EOS may be different from $\hat{p} = w
\hat{\rho}$. Fluctuations of the ratio $\frac{\Gamma}{3H}$ can be
regarded as fluctuations of the EOS which give rise to pressure
perturbations $\hat{p} = w \hat{\rho} + \hat{w}\rho$, where
$\hat{w} = \left(\frac{\Gamma}{3 H}\right)^{\hat{}}$. Thus, the
equivalent two-component picture represents a simple method to
construct variations of the cosmic EOS while keeping $w_M =0$ and
$w_X =$ const. On this basis one may reconsider fluid models of
the cosmic substratum which occasionally are abandoned since a
purely adiabatic treatment does not correctly reflect the
observational situation. Taking into account an internal structure
of the medium, which is equivalent to go beyond the limits of an
adiabatic analysis, will generally result in a different picture.

The internal structure, modelled here through a fluctuating decay
rate, will also influence the large scale perturbation behavior.
Large scale perturbations are conveniently described by the
quantity $\zeta$ which represents a curvature perturbation on
hypersurfaces of constant energy density \cite{Bardeenetal83}. On
large perturbation scales $\zeta$ obeys the equation (cf.
Refs.~\cite{Lyth} - \cite{Wandsetal00})
\begin{equation}\label{dotzeta}
\dot{\zeta} = - H \frac{\hat{p} -
\frac{\dot{p}}{\dot{\rho}}\hat{\rho}}{\rho + p}\ .
\end{equation}
For adiabatic perturbations $\hat{p} =
\frac{\dot{p}}{\dot{\rho}}\hat{\rho}$ the quantity $\zeta$ is
conserved. As soon as there are non-adiabatic perturbations,
$\zeta$ will no longer be constant. For our model with $w > -1$
(and, for simplicity, time independent perturbations $\hat{w} =
\left(\frac{\Gamma}{3 H}\right)^{\hat{}}$\,) we find
\begin{equation}\label{zeta=}
\zeta = \zeta_{i} - \frac{1}{1+w}\left(\frac{\Gamma}{3
H}\right)^{\hat{}}\ln\left(\frac{a}{a_{i}}\right) \ .
\end{equation}
A fluctuating fractional decay rate gives rise to a change in the
curvature perturbation which is logarithmic in the scale factor.
The subscript $i$ denotes some initial value. The circumstance
that $\zeta$ is time dependent has consequences for the time
dependence of the (gauge-invariant) gravitational potential for
which we obtain
\begin{equation}
\Phi = - 3\, \frac{1+w}{5+3w}\,\zeta_{i} +
\frac{9}{2}\frac{1}{5+3w}\left(\frac{\Gamma}{3\,H}\right)^{\hat{}}\left[\ln
{\left(\frac{a}{a_i}\right)} - \frac{3}{5+3w}\right] \
.\label{Phisol}
\end{equation}
While the first constant term on the right-hand side is a standard
result for adiabatic perturbations \cite{LL}, the fluctuating
decay rate introduces a logarithmic dependence on the scale
factor. The time dependence of $\Phi$,
\begin{equation}
\dot{\Phi} =
\frac{9}{2}\frac{1}{5+3w}\left(\frac{\Gamma}{\Theta}\right)^{\hat{}}\,H
\ ,\label{dotPhi}
\end{equation}
determines the ISW. Different from the pure adiabatic case where
$\dot{\Phi} = 0$ there is an ISW in this case. It is interesting
to compare the result (\ref{Phisol}) with the corresponding
expression of the $\Lambda$CDM model. In the long time limit one
finds for the latter $\Phi_{\small \Lambda{\rm CDM}} \approx -
\frac{3}{4}\zeta_{i}\frac{\rho_{M_i}}{\rho_{\Lambda}}
\frac{a_{i}}{a} $ where $\rho_{M_i}$ is some initial value and
$\rho_{\Lambda} = {\rm const}$. This means, $\Phi_{\small
\Lambda{\rm CDM}}$ shows a stronger time dependence than $\Phi$ in
(\ref{Phisol}). Apparently, our interacting DE model predicts a
smaller ISW compared with the non-interacting $\Lambda$CDM model.
A smaller ISW is equivalent to a suppression (compared with the
prediction of the $\Lambda$CDM model) of the lowest multipoles in
the anisotropy spectrum of the cosmic microwave background.  Such
kind of suppression is required by the data and different
mechanisms have been invented to account for this phenomenon
\cite{Contaldi,BoFeng,Bastero,Tsujikawa}. Here we argue that it
may well be the consequence of an interaction in the dark sector.

\section{Summary}

Interacting DE models are more general than non-interacting ones.
A coupling between DE and DM enriches the cosmological dynamics.
It reveals potential degeneracies in interpreting the
observational data and provides a method to refine the equation of
state of the cosmic substratum. An interaction is particularly
useful to address the coincidence problem. It also opens new
perspectives on holographic DE and on the big rip scenario.
Furthermore, it introduces a non-adiabatic feature into the
perturbation dynamics with consequences for the effective sound
speed and the time dependence of the gravitational potential. The
interaction should be detectable by suitably refined luminosity
distance measurements and by certain features in the anisotropy
spectrum of the cosmic microwave background.

\end{document}